\documentclass[epjCONF, onecolumn]{svjour}
\usepackage{graphics}
\usepackage[square, sort, numbers]{natbib}

\usepackage[varg]{txfonts} % Times fonts
\usepackage[latin1]{inputenc}
\session-title{Conference Title, to be filled}
\begin{document}
\title{Solar-like pulsating stars as distance indicators: G-K giants in the CoRoT and {\it Kepler} fields}
\author{Andrea Miglio\inst{1}\fnmsep\thanks{\email{a.miglio@bham.ac.uk}} \and Thierry Morel\inst{2} \and   Mauro Barbieri\inst{3} \and Beno\^it Mosser\inst{4}, L\'eo Girardi\inst{5} \and Josefina Montalb{\'a}n\inst{2} \and  Marica Valentini\inst{4}}
%ºb
\institute{School of Physics and Astronomy, University of Birmingham, UK \and Institut d'Astrophysique et de G\'eophysique, Li\`ege, Belgium  \and Observatoire de la C{\^o}te d'Azur, Nice, France \and LESIA, Observatoire de Paris, France \and INAF, Osservatorio Astronomico di Padova }
\abstract{
The detection of radial and non-radial solar-like oscillations in thousands of G-K giants with CoRoT and {\it Kepler} is paving the road for detailed studies of stellar populations in the Galaxy. The available average seismic constraints allow a precise and largely model-independent determination of stellar radii (hence distances) and masses. We here briefly report on the distance determination of thousands of giants in the CoRoT and {\it Kepler} fields of view.} %end of abstract
\maketitle
%
%\section{Distance}
%\label{intro}
Thanks to the interpretation of solar-like oscillation spectra detected by CoRoT and {\it Kepler} \cite{DeRidder09, Mosser2010, Bedding10, Hekker2011b}, we can determine the mass and radius of thousands of stars belonging to the composite population of the Milky Way's disk. These innovative constraints allow precise estimates of distances and ages for giants, and will inform studies of galactic formation and evolution with observational constraints which were not available prior to asteroseismology (see e.g. \cite{Miglio2009, Miglio2012, Freeman2012, Chiappini2012} and references therein). We here briefly report on the distance determination of giants in the CoRoT and {\it Kepler} fields of view. A detailed description of the data, method, and results will be presented in a forthcoming paper.

As a first step we determine stellar radii by combining the available seismic parameters $\nu_{\rm max}$ and $\Delta\nu$ with effective temperatures $T_{\rm eff}$. The latter are determined using 2MASS photometry and the colour-$T_{\rm eff}$ calibrations by \cite{Alonso1999}. 
We then compute luminosities  $L$ from $R$ and $T_{\rm eff}$, and distances combining $L$ with de-reddened apparent 2MASS Ks magnitudes and bolometric corrections from \cite{Girardi05}.  Distance-dependent extinction from \cite{Drimmel2003} is considered when determining the distance and $T_{\rm eff}$. We estimate the uncertainty on the distances to be $\sim 10-15\%$. This value can be further reduced when spectroscopic constraints will be available, and additional empirical tests of the scaling relations will be performed. 

\begin{figure}
\centering
% Use the relevant command for your figure-insertion program
% to insert the figure file.
% For example, with the option graphics use
\resizebox{0.76\columnwidth}{!}{%
  \includegraphics{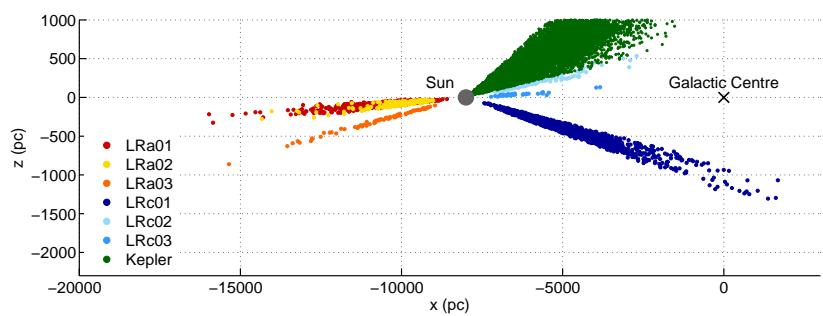} }
\caption{Solar-like oscillating G-K giants observed in several CoRoT fields of view and by {\it Kepler}: projection on the $x-z$ plane.}
\label{lab:fig1}       % Give a unique label
\end{figure}

\begin{figure}[ht]
\begin{minipage}[t]{0.46\linewidth}
%\centering
\resizebox{1.1\columnwidth}{!}{%
\includegraphics{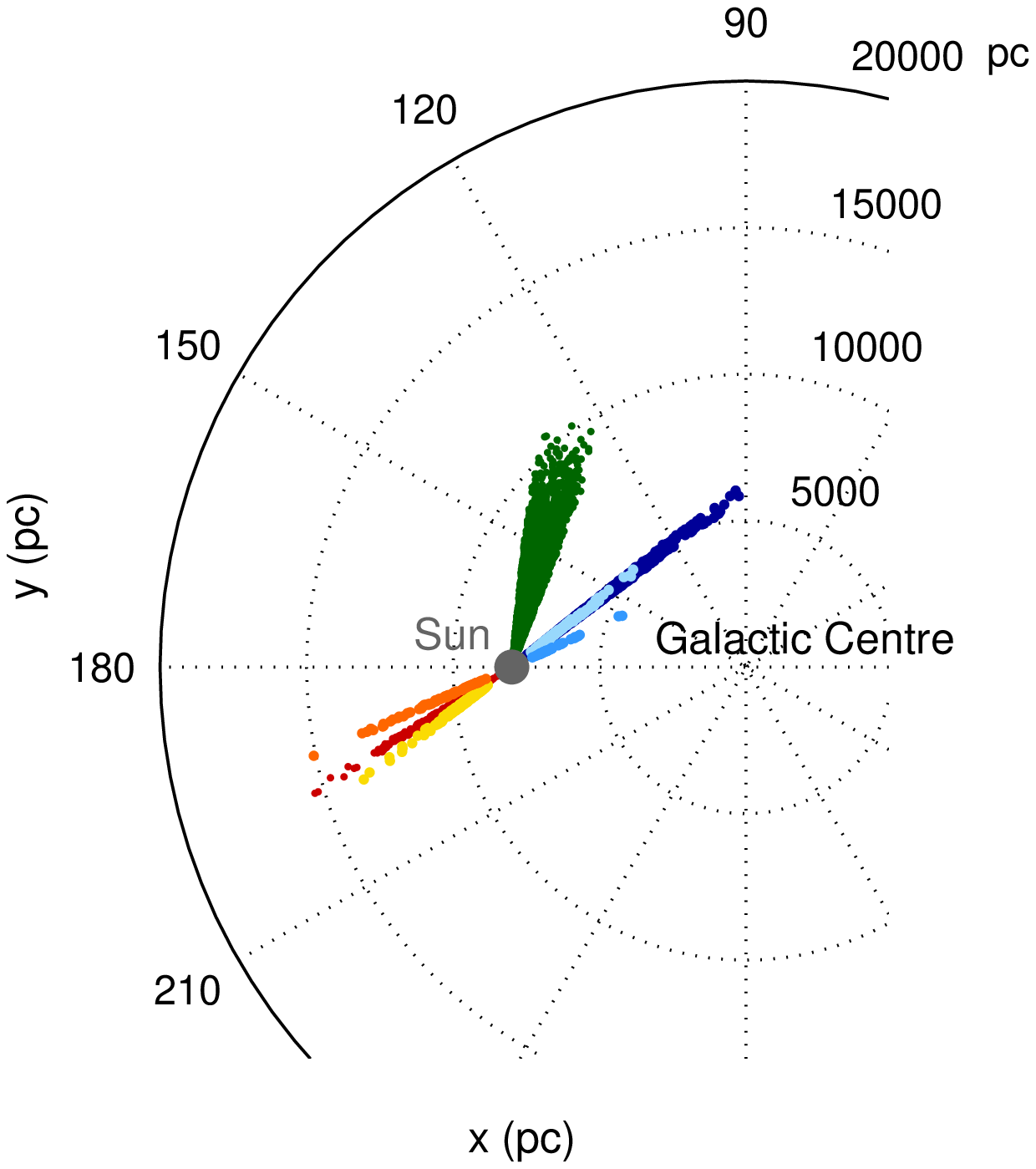}}
\centering
\caption{Projection on the Galactic plane of the stars shown in Fig. \ref{lab:fig1}.}
\label{lab:fig2}
\end{minipage}
\hspace{0.5cm}
\begin{minipage}[t]{0.46\linewidth}
\centering
\resizebox{1.02\columnwidth}{!}{%
\includegraphics{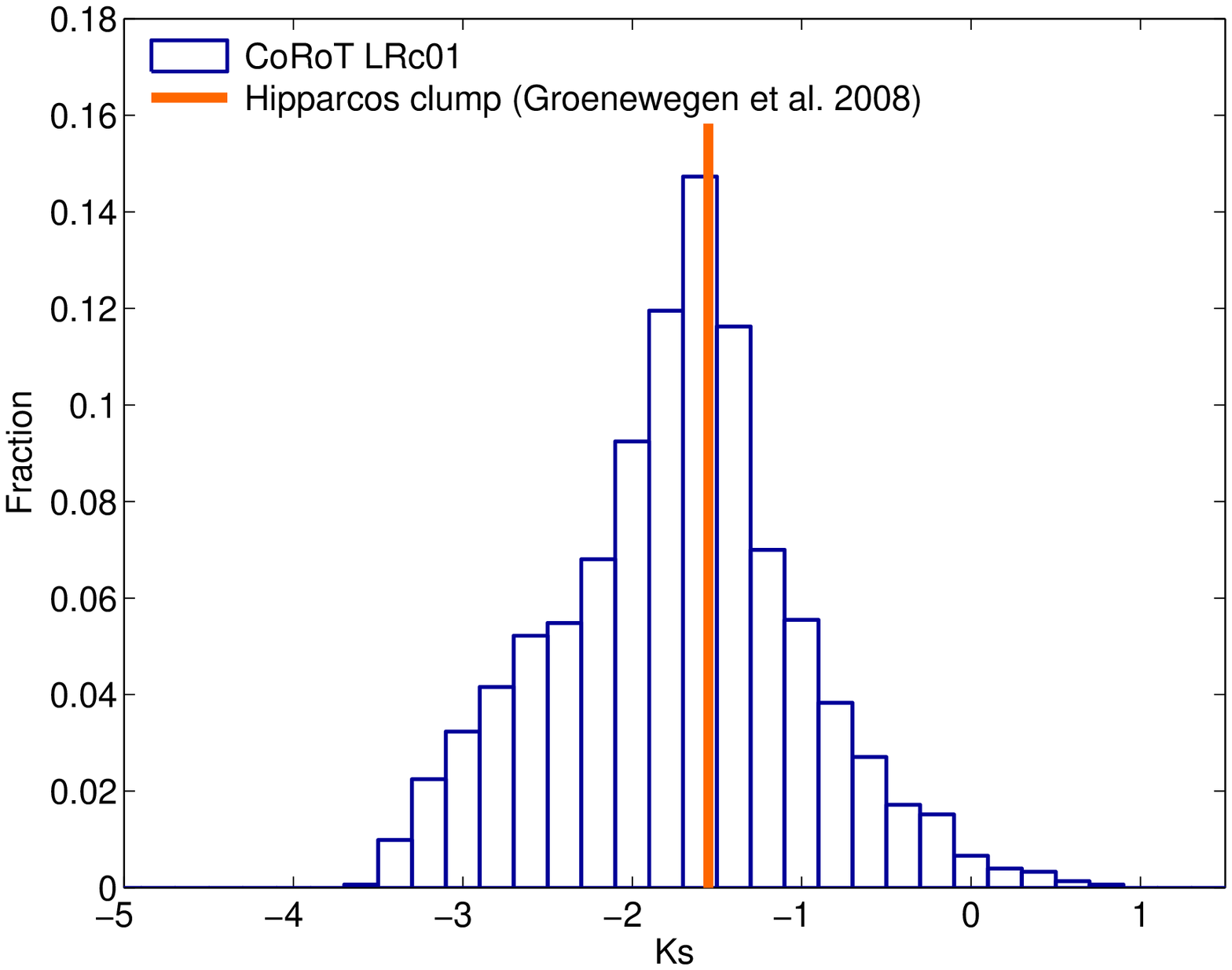}}
\caption{Distribution of the derived absolute Ks magnitude of solar-like pulsating giants in CoRoT LRc01. The vertical line shows the absolute magnitude of the Hipparcos red clump \cite{Groenewegen08}.}
\label{lab:fig3}
\end{minipage}
\end{figure}

We apply this procedure to pulsating giants observed by CoRoT in several observational runs, and to giants in the public {\it Kepler} data \cite{Hekker2011b}. The location in the Galaxy of $\sim 2500$ CoRoT and  $\sim10000$ $Kepler$ targets is shown in Fig. \ref{lab:fig1} and Fig. \ref{lab:fig2}. 
As presented in Fig. \ref{lab:fig3}, the peak in the distribution of the absolute Ks magnitudes of giants (e.g. in CoRoT's LRc01 field) is in remarkable agreement with the absolute Ks magnitude of Hipparcos  red-clump giants \cite{Groenewegen08}. 

Studies are currently underway to combine distances with spectroscopic constraints, as well as with asteroseismic estimates of the mass (hence age) of these targets, leading to a detailed characterisation of populations of giants in different regions of the Milky Way.

%
%\begin{figure}
%
%\begin{minipage}{0.4\linewidth}
%% Use the relevant command for your figure-insertion program
%% to insert the figure file.
%% For example, with the option graphics use
%\resizebox{1\columnwidth}{!}{%
%  \includegraphics{./figure/polar.eps} }
%  \caption{}
%\label{lab:fig2}
%\end{minipage}
%
%\begin{minipage}{0.4\linewidth}
%% Use the relevant command for your figure-insertion program
%% to insert the figure file.
%% For example, with the option graphics use
%\resizebox{\columnwidth}{!}{%
%  \includegraphics{./figure/hipp.eps} }
%  \caption{}
%\label{lab:fig3}
%\end{minipage}
%
%\end{figure}

%
%\begin{minipage}[r]{0.33\linewidth}
%\vspace*{-7.cm}
%\caption{Lithium data for field evolved stars from the sample by Charbonnel et al. (in prep.) that are segregated according to their mass (left and right panels include respectively sample stars with masses lower and higher than 2~M$_{\odot}$; Li detections and upper limits are shown as circles and triangles respectively). 
%	  Theoretical lithium evolution is shown from the ZAMS up to the end of the early-AGB. 
%	  Various lines correspond to predictions for stellar models of different masses computed without or with rotation as indicated, and with thermohaline mixing in all cases.}
%\label{fig2}       % Give a unique label
%\end{minipage}\hfill
%\label{fig:1}       % Give a unique label
%\end{figure}

\begin{acknowledgement}
AM acknowledges financial support from the organisers and from the School of Physics and Astronomy, University of Birmingham.
\end{acknowledgement}

\bibliographystyle{epj}
\bibliography{../../andrea_m.bib}

%\begin{thebibliography}{}
%% and use \bibitem to create references.
%\bibitem{RefJ}
%% Format for Journal Reference
%Author, Journal \textbf{Volume}, (year) page numbers
%% Format for books
%\bibitem{RefB}
%Author, \textit{Book title} (Publisher, place year) page numbers
%% etc
%\end{thebibliography}

\end{document}